\documentclass[12pt,twoside]{article}
\usepackage{amssymb}
\usepackage{amsmath}
\begin{document}
\setcounter{page}{1}
\newtheorem{t1}{Theorem}[section]
\newtheorem{d1}{Definition}[section]
\newtheorem{c1}{Corollary}[section]
\newtheorem{l1}{Lemma}[section]
\newtheorem{r1}{Remark}[section] 
\newcommand{\cA}{{\cal A}}
\newcommand{\cB}{{\cal B}}
\newcommand{\cC}{{\cal C}}
\newcommand{\cD}{{\cal D}}
\newcommand{\cE}{{\cal E}}
\newcommand{\cF}{{\cal F}}
\newcommand{\cG}{{\cal G}}
\newcommand{\cH}{{\cal H}}
\newcommand{\cI}{{\cal I}}
\newcommand{\cJ}{{\cal J}}
\newcommand{\cK}{{\cal K}}
\newcommand{\cL}{{\cal L}}
\newcommand{\cM}{{\cal M}}
\newcommand{\cN}{{\cal N}}
\newcommand{\cO}{{\cal O}}
\newcommand{\cP}{{\cal P}}
\newcommand{\cQ}{{\cal Q}}
\newcommand{\cR}{{\cal R}}
\newcommand{\cS}{{\cal S}}
\newcommand{\cT}{{\cal T}}
\newcommand{\cU}{{\cal U}}
\newcommand{\cV}{{\cal V}}
\newcommand{\cX}{{\cal X}}
\newcommand{\cW}{{\cal W}}
\newcommand{\cY}{{\cal Y}}
\newcommand{\cZ}{{\cal Z}}
\def\cl{\centerline}
\def\bd{\begin{description}}
\def\be{\begin{enumerate}}
\def\ben{\begin{equation}}
\def\benn{\begin{equation*}}
\def\een{\end{equation}}
\def\eenn{\end{equation*}}
\def\benr{\begin{eqnarray}}
\def\eenr{\end{eqnarray}}
\def\benrr{\begin{eqnarray*}}
\def\eenrr{\end{eqnarray*}}
\def\ed{\end{description}}
\def\ee{\end{enumerate}} 
\def\al{\alpha}
\def\b{\beta}
\def\bR{\bar\R}
\def\bc{\begin{center}}
\def\ec{\end{center}}
\def\d{\dot}
\def\D{\Delta}
\def\del{\delta}
\def\ep{\epsilon}
\def\g{\gamma}
\def\G{\Gamma}
\def\h{\hat}
\def\iny{\infty}
\def\La{\Longrightarrow}
\def\la{\lambda}
\def\m{\mu}
\def\n{\nu}
\def\noi{\noindent}
\def\Om{\Omega}
\def\om{\omega}
\def\p{\psi}
\def\pr{\prime}
\def\r{\ref}
\def\R{{\bf R}}
\def\ra{\rightarrow}
\def\s{\sum_{i=1}^n}
\def\si{\sigma}
\def\Si{\Sigma}
\def\t{\tau}
\def\th{\theta}
\def\Th{\Theta}
\def\vep{\varepsilon}
\def\vp{\varphi}
\def\pa{\partial}
\def\un{\underline}
\def\ov{\overline}
\def\fr{\frac}
\def\sq{\sqrt}
\def\WW{\begin{stack}{\circle \\ W}\end{stack}}
\def\ww{\begin{stack}{\circle \\ w}\end{stack}}
\def\st{\stackrel}
\def\Ra{\Rightarrow}
\def\R{{\mathbb R}}
\def\bi{\begin{itemize}}
\def\ei{\end{itemize}}
\def\i{\item}
\def\bt{\begin{tabular}}
\def\et{\end{tabular}}
\def\lf{\leftarrow}
\def\nn{\nonumber}
\def\va{\vartheta}
\def\wh{\widehat}
\def\vs{\vspace}
\def\Lam{\Lambda} 
\def\sm{\setminus}
\def\ba{\begin{array}}
\def\ea{\end{array}}
\def\ds{\displaystyle}
\def\lan{\langle}
\def\ran{\rangle}

\baselineskip 15truept

\bc
{\Large {\bf Quantum correlations in successive spin measurements}} \\

{\bf Ali Ahanj\footnote{Electronic address: ahanj@physics.unipune.ernet.in} and Pramod Joag\footnote{Electronic address : pramod@physics.unipune.ernet.in}} \\
Department of Physics,  University of Pune,  Pune - 411007, India
\ec

In this paper we present a new approach for testing QM against the realism aspect of  hidden variable theory (HVT). We consider successive measurements of non-commuting operators on a input spin $s$ state. The key point is that, although these operators are non-commuting, they act on different states so that the joint probabilities for the outputs of successive measurements are well defined. We show that, in this scenario HVT leads to Bell type inequalities for the correlation between the outputs of successive measurements. We account for the maximum violation of these inequalities by quantum correlations by varying spin value and the number of successive measurements. Our approach can be used to obtain a measure of the deviation of QM from realism say in terms of the amount of information needed to be transferred between successive measurements in order to classically simulate the quantum correlations.


\section{INTRODUCTION}

Quantum Mechanics (QM) is known to be nonlocal or nonrealistic and contextual [1]. All theories and experiments to test these aspects of QM are based on the multipartite quantum systems in entangled states. Although this scenario is inevitable for the tests of nonlocality, it is not obligatory for testing realism and contextuality. In this paper we propose and analyse a particular scenario to account for the deviations of QM from realism, which involves correlations in the outputs of successive measurements of noncommuting operators on a spin $s$ state.\\

The successive measurement correlations have been used previosly in the context of non-local correlations by Popescu[2], in order to analyse a class of Werner states which are entangled but do not break (bipartite) Bell type inequality. Although local HVT can simulate the quantum correlations between the outputs of single ideal measurement on each part of the system, it fails to simulate the correlations of the second measurements on each part. Leggett and Garg have used consecutive measurements to challenge the applicability of QM to macroscopic phenomena[3]. Finally there is a large literature on the problem of information of a quantum state that can be obtained by measuring the same operator successively on a single system. The research in this area is elegently summerized in[4].\\

The paper is organized as follows. In Section 2 we describe the basic scenario in detail. Section 3 formulates the implications of hidden variable theory (HVT) for this scenario in terms of Bell type inequalities. Section 4 evaluates these inequalities for mixed spin $s$ input states for two and three successive measurements for various spin values. Section 5 deals with $n$ successive measurements on spin $s$ system. In section 6 we give a protocol to simulate the correlations between $n$ successive measurements on a spin 1/2 system. Finally we conclude with  summary and comments in Section 7. Mathematical details are relegated to Appendices A and B. 

\section{BASIC SCENARIO}

Consider the following sequence of measurements. A quantum particle with spin $s$ prepared in the initial state $\rho_0$ is sent through a string of Stern-Gerlach (SG) measurements for the spin components along the directions given by the unit vectors $\h a_1, \h a_2, \h a_3, \cdots, \h a_n$. Each measurement has $2s+1$ possible outcomes. For the $i$-th measurement, we denote these outcomes (eigenvalues) by $\al_i \in \{ s, s-1, \cdots, -s\}$. We denote by $\langle \al_i \rangle$ the quantum mechanical (ensemble) average $\langle \vec{s} \cdot \h a_i \rangle,$ by $ \langle \al_i  \al_j \rangle$ the average $\langle (\vec{s} \cdot \h a_i) (\vec{s} \cdot \h a_j) \rangle$ etc.

Each of the $(2s+1)^n$ possible outcomes after $n$-th measurement corresponds to a particular combination of the results of the previous measurements and the probability of these outcomes is the joint probability for such combinations. Note that in this case these joint probabilities are well defined, even if $\vec{s} \cdot \h a_i (i = 1, 2, \cdots, n)$ do not commute, because each of these operators act on different states [5,6,7].  We emphasize that this is the joint probability for the results of  $n$ actual measurements and not a joint distribution for hypothetical simultaneous values of $n$ noncommuting observables. Moreover, various subbeams in this expriment are separated without any overlap on recombination between them. We further assume that, between two successive measurements, the spin state does not change with time i.e. $\vec{s}$ commutes with the interaction Hamiltonian, if any. Also, throughout the string of measurements, no component is blocked.

\section{IMPLICATIONS OF HVT}

HVT assumes that in every possible state of the system, all observables have well defined (sharp) values [8]. On the measurement of an observable in a given state, the value possessed by the observable in that state (and no other value) results.
To gain compatibility with QM and the experiments, a set of `hidden' variables is introduced which is denoted collectively by $\la$. For given $\la$, the values of all observables are specified as the values of appropriate real valued functions defined over the domain $\Lambda$ of possible values of hidden variables. For the spin observable $\vec{s} \cdot \h a$, we denote the value of $\vec{s} \cdot \hat a$ in the QM (spin) state $| \psi \rangle$ by $\al$. Considered as a function $\al : \Lambda \ra I\!\!R$ we represent the value of $\vec{s} \cdot \h a$ when the hidden variables have the value $\la$ by $\al(\la)$. More generally, we may require that a value of $\la$ gives the probability density $p(\al | \la)$ over the values of $\al$ rather than specifying the value of $\al$ (stochastic HVT). We denote the probability density function for the hidden variables in the state $| \psi \rangle$ by $\rho_\psi$.( $\rho_\psi(\la) d\la$ measures the probability that the collective hidden variable lies in the range $\la$ to $\la + d \la$). Then the average value of $\vec{s} \cdot \h a$ in the state $|\psi \rangle$ is 
$$ \langle \al \rangle = \int_\Lambda \al(\la) \rho_\psi(\la) d\la \eqno{(3.1)}$$
where the integration is over $\Lambda$ defined above. In the general case (SHVT)
$$ \langle \al \rangle = \int_\Lambda \al p(\al | \la) \rho_\psi(\la) d\la \eqno{(3.2)}$$ 
We now analyse the consequences of SHVT for our scenario. In general, the outputs of $k$th and $l$th experiments may be correlated so that, $$ p(\al_i, \h a_k \& \al_j, \h a_\ell)\neq p(\al_i; \h a_k)p(\al_j; \h a_l)\eqno{(3.3)}$$
However, in SHVT we suppose that these correlations have a common cause represented by a stochastic hidden variable $\lambda$ so that
$$ p(\al_i, \h a_k \& \al_j, \h a_\ell | \la) = p(\al_i, \h a_k|\la) p(\al_j, \h a_l | \la)\eqno{(3.4)}$$

This is the crucial equation expressing the fundamental implication of SHVT to the successive measurement scenario. We now obtain the Bell type inequalities from equation (3.4) which can be compared with QM.
Here we assume that in HVT all probabilities corresponding to outputs of measurements account for the possible changes in the values of the observable being measured, (due to the interaction of the measuring device and the system), occurring in the previous measurements.\\
Now consider (dropping $\h a_k, \h a_\ell)$
$$ \langle \al_i , \al_j \rangle = \int \rho(\la) E(\al_i, \al_j, \la) d\la \eqno{(3.5)}$$ 
where 
\begin{eqnarray*}
 E(\al_i, \al_j, \la)& =& \sum_{\al_i, \al_j} \al_i \al_j p(\al_i, \al_j | \la)=  \sum_{\al_i} \al_i p(\al_i | \la) \sum_{\al_j} \al_j p(\al_j | \la)\\
& =& E(\al_i, \la) E(\al_j, \la)~~~~~~~~~~~~~~~~~~~~~~~~~~~~~~~~~~~~~~~~~~~~~~~~~~~~~~~~~~~~~~~~~~(3.6)
\end{eqnarray*}
Now let us consider the case of two successive measurements, with options $\h a_1, \h a_1'$ and $\h a_2, \h a_2'$ respectively for measuring spin components. In each run of the experiment, a random choice between $\{ \h a_1, \h a_1'\}$ and $\{ \h a_2, \h a_2'\}$ is made. Define $ \th_i$ ($ i=1,1' $) to be the angle between $\h a_i$ and $\h a_{0} $, $ \th_{ij}$  ($ i=1,1' $ and $ j =2, 2' $) is the angle between $\h a_j$ and $\h a_i$. Using condition (3.6) and the result [9,10] 
$$ - 2s^2 \le xy + xy' + x'y - x'y' \le 2s^2,~~~~~ 
 x, y, x', y' \in [-s, + s]. $$ 
We obtain 
$$ -2s^2 \le E(\al_1, \al_2, \la) + E(\al_1, \al_2', \la) + E(\al_1', \al_2, \la) - E(\al_1', \al_2', \la) \le 2s^2.$$
Multipling by $\rho(\la) d\la$ and integrating over $\Lambda$, we get Bell inequality for two successive measurement outputs: 
$$ |\lan BI \ran| =\frac{1}{2} |\lan \al_1 \al_2 \ran + \lan \al_1 \al_2'\ran + \lan \al_1' \al_2\ran - \lan \al_1' \al_2'\ran | \le s^2 \eqno{(3.7)}$$ 
Similarly using 
$$ - 2s^3 \le xyz' + xy'z + x'yz - x'y'z' \le 2s^3 ,~~~ 
 x, y, z, x', y', z' \in [-s, s]$$ 
and 
$$ E(\al_i, \al_j, \al_k, \la) = E(\al_i, \la) E(\al_j, \la) E(\al_k, \la)$$ We can prove Mermin-Klyshko Inequality (MKI) for three successive measurements, 
$$ |\lan MKI\ran| = \frac{1}{2}| \lan \al_1 \al_2 \al_3'\ran + \lan \al_1 \al_2' \al_3 \ran + \lan \al_1' \al_2 \al_3 \ran - \lan \al_1' \al_2' \al_3'\ran | \le s^3 . \eqno{(3.8)}$$ 

Let $|  \lan MKI'\ran | \le s^3$. 
$|\lan MKI'\ran| $ is obtained by exchanging primes with nonprimes and vice-versa in MKI

$$ | \lan SI \ran| = | \lan MKI \ran +  \lan MKI'\ran| \le| \lan MKI \ran| + | \lan MKI'\ran|  \le 2s^3 \eqno{(3.9)}$$ 
This is the svetlichny  inequality (SI). [11,12,13]

For $n$ successive measurements on spin $s$ system, we define the MK polynomials recursively as follows:\\
$$ M_1 = \al_1, M_1' = \al_1' \eqno{(3.10)} $$ 
$$ M_n = \fr{1}{2} M_{n-1} (\al_1 + \al_1') + \fr{1}{2} M_{n-1}' (\al_1 - \al_1') \eqno{(3.11)}$$ 

where $M_n'$ are obtained from $M_n$ by exchanging all primed and non-primed $\al$'s. 
The recursive relation(3.11) gives, for all $ 1\leq k\leq n-1$ [13]:
$$ M_n = \fr{1}{2} M_{n-k} (M_k + M_k') + \fr{1}{2} M_{n-k}' (M_k - M_k') \eqno{(3.12)}$$ 

In particular, we have 
$$ M_2 = BI  = \fr{1}{2} (\al_1 \al_2 + \al_1'\al_2 + \al_1\al_2' - \al_1' \al_2') \eqno{(3.7)}$$ 

$$ M_3 = MKI = \fr{1}{2}(\al_1\al_2 \al_3' + \al_1\al_2'\al_3 + \al_1' \al_2\al_3 -\al_1'\al_2'\al_3')\eqno{(3.8)} $$

We show that in HVT  $$ | \lan M_n \ran | \le s .\eqno (3.13)\\$$ 

\noi First note that (3.13) is true for $n = 1, 2, 3$ (equations (3.7), (3.8)). Suppose it is true for $n = k$ i.e. $Max |\lan M_k\ran | =s$. Now 
$$ | \lan M_{k+1} \ran | = \fr{1}{2} | \lan M_k \al_{k+1}\ran + \lan M_k \al_{k+1}' \ran + \lan M_k' \al_{k+1} \ran - \lan M_k' \al_{k+1}' \ran |$$ 
Since HVT applies here we can use (3.4) to get 
$$ |\lan M_{k+1}\ran| = \fr{1}{2} |\lan M_k\ran (\lan \al_{k+1} \ran + \lan \al_{k+1}'\ran ) + \lan M_k'\ran (\lan \al_{k+1}\ran - \lan \al_{k+1}' \ran )|$$ 
This implies, by induction hypothesis, that 
$$ \max | \lan M_{k+1} \ran | = \max | \lan M_k\ran| = s$$

\section{MIXED INPUT STATE FOR ARBITRARY SPIN} 

\subsection{Two successive measurements (BI) } 

We first deal with the case when input state is a mixed state whose eigenstates coincide with those of $\vec{s} \cdot \h a_0$ for some $\h a_0$ whose eigenvalues we denote by $\al_0\in\{ -s, \cdots s\}$. For spin 1/2 this is the most general mixed state because given any density operator $\rho_0$ for spin 1/2 (corresponding to some point within the Bloch sphere), we can find an $\h a_0$ such that the eigenstates of $\vec{s} \cdot \h a_0$ and $\rho_0$ coincide. However, for $s > 1/2$, our choice forms a restricted class of mixed states. We note that these are the only states accessible via SG expriments. Thus we have 
$$ \rho_0 = \sum_{\al_0} p_{\al_0} | \vec{s} \cdot \h a_0 , \al_0 \ran \lan \vec{s} \cdot \h a_0, \al_0 | ;~~~~~~\left( \sum_{\al_0} p_{\al_0}=1 \right)  \eqno{(4.1)}$$ 
After the first measurement along $\h a_1$, the resulting state of the system is 
$$ \rho_1 = \sum_{\al_1} M^\dagger_{\al_1} \rho_0 M_{\al_1} \eqno{(4.2)}$$ 
$$ M^\dagger_{\al_1} = M_{\al_1} = | \vec{s} \cdot \h a_1, \al_1 \ran \lan \vec{s} \cdot \h a_1, \al_1 | .$$ 
Now 
$$
\lan \al_1 \al_2 \ran  =  Tr(\rho_1 \vec{s} \cdot \h a_1 \vec{s} \cdot \h a_2) 
 =  \sum_{\al_0 \al_1 \al_2} p_{\al_0} \al_1 \al_2 | \lan \vec{s} \cdot \h a_0, \al_0 | \vec{s} \cdot \h a_1, \al_1 \ran |^2 | \lan \vec{s} \cdot \h a_1, \al_1 | \vec{s} \cdot \h a_2, \al_2 \ran |^2  \eqno{(4.3)}$$
By equation (A.12), we get 
$$ \lan \al_1 \al_2 \ran = \fr{1}{2} \cos \th_{12} [A \cos^2 \th_{1} + B] \eqno{(4.4)}$$
where 
$$A  =  3 \chi - s(s+1) ,~~~B  =  s(s+1) - \chi,~~~\chi  =  \sum^{+s}_{\al_0= - s} \al^2_0 p_{\al_0}.$$
This leads to the following expression for the Bell inequality: 
$$BI  =  \fr{1}{4} (A \cos^2 \th_{1} + B) (\cos \th_{12} + \cos \th_{12'})
 + \fr{1}{4} (A \cos^2 \th_{1'} + B) (\cos \th_{1'2} - \cos \th_{1'2'}) \eqno{(4.5)}$$
We introduce ${\ds \eta = \fr{|BI|}{s^2}}$. If $\eta > 1$ two successive measurements violate HVT.
For a given $\rho_0, \eta$ is maximized for 
$ \th_1 + \th_1' = \pi ;~~ \th_2 = \fr{\pi}{2}; ~~~ \th_2' = 0$ 
This gives 
$$ \eta = \left( \fr{1}{2s^2}\right) [ (\sin \th_1 + \cos \th_1) (A \cos^2 \th_1 + B)] \eqno{(4.6)}$$
${\ds \fr{\pa \eta}{\pa \th_1} = 0}$ implies 
$$ B \tan^3 \th_1 + (2A-B) \tan^2 \th_1+ (3A + B) \tan \th_1 - (A+B) = 0 . \eqno{(4.7)}$$ 
Real roots of this equation give values of $\th_1$ for which $\eta$ is maximum. The maximum value $\eta$ is evaluated at these $\th_1$.\\

 We find that for $s=\frac{1}{2} $, $ \chi = 1/4$ for all  $\rho_0$, so  $\eta_{\max} = \sq 2$. Thus all possible spin 1/2 states break BI for two successive measurements. This can be compared with the two particle scenario where only the entangled pure states break BI while not all entangled mixed states break it[14].\\ 

For spin 1 all states which do not have any contribution of $s_z = 0$ eigenstate break BI. In this case $\chi = 1$ for all $\rho_0$ and $ \eta_{\max} (s = 1) \cong 1.2112.$
When the $s_0 = 0$ state contributes, all $\rho_0 s$ with $0 \le p(\al_0 = 0) < 0.23$ and $0.67 < p(\al_0 = 0) \le 1$ break BI, while others satisfy it. Notice that, when $p(\al_0 = 0) = 1$ i.e. $ \rho_0 = | \vec{s} \cdot \hat{a}_0, 0 \ran \lan  \vec{s} \cdot \hat{a}_0, 0 | $ we have  violation of  BI given by  $ \eta_{\max}(s = 1) =  1.143 $. \\

For all $s > 1$  BI is broken when the states $s_z = \pm s$ contribute significantly as can be seen in table 1 (we introduce $\xi = \chi/s^2$). \\
\newpage

\bc
{\bf Table 1} \\
\vspace{.1in} 

\bt{||c|c||c|c||c|c||}
\hline 
$s$ & $\xi$ & $s$ & $\xi$ & $s$ & $\xi$ \\
\hline 
$\fr{1}{2}$ & $0 \le \xi \le 1$ & $\fr{5}{2}$ & $0.847 \le \xi \le 1$ & $\fr{9}{2}$ & $0.858 \le \xi \le 1$\\
\hline 
1 & $ 0 \le \xi \le 0.33$ and  $0.77 \le \xi \le 1$ & 3 & $0.851 \le \xi \le 1$  & 5 & $0.859 \le \xi \le 1$\\
\hline 
$\fr{3}{2} $ & $0.824 \le \xi \le 1 $  & $\fr{7}{2}$ & $0.854 \le \xi \le 1$ & $\fr{11}{2}$ & $0.860 \le \xi \le 1$  \\
\hline 
2 & $0.84 \le \xi \le 1$  & 4 & $0.856 \le \xi \le 1$  
& 6 & $0.862 \le \xi \le 1$ \\
\hline 
& & & & $\iny$ & $0.87 \le \xi \le 1$  \\
\hline 
\et

\vspace{.2in}

The range $\xi$ for the violation  of BI  \\
\ec

Note that $\eta_{\max} $ is realized for states of the form 
$$ \rho^{\max}_0 = p_s | \vec{s} \cdot \h a_0, s \ran \lan \vec{s} \cdot \h a_0,s| + p_{-s}|  \vec{s} \cdot \h a_0 , - s\ran \lan \vec{s} \cdot \h a_0, -s | \eqno{(4.8)}$$ 
$$p_s + p_{-s} = 1 $$ 
From Table (1), it is clear that when  $\chi = s^2 (\xi=1)$ maximum violation of BI is obtained. 

Next we can also see that, for $s < 15$, except $s=1$ when $\rho_0$ does not have any contribution from $\al_0 = \pm s$ states, it satisfies BI. Consider 
$$ 1 \ge \xi = (p_s + p_{-s}) + (p_{s-1} + p_{-s+1}) \fr{(s-1)^2}{s^2} + \cdots \ge X$$
which is the required condition on $\xi$ for breaking of the BI, where $X \le \xi \le 1$ ($X$ varies between 0.82 and 0.87 for $s \ge 1$ as shown in Table 1). When $p_s = p_{-s} = 0$  we must have 
$$ (p_{s-1} + p_{-s+1}) \fr{(s-1)^2}{s^2} + (p_{s-2} + p_{-s+2}) \fr{(s-2)^2}{s^2} + \cdots \ge X.$$
But $LHS < (\frac{s-1}{s})^2$ which is less than $X$ for $s < 15$ as seen from the Table 1. So for $s < 15$, maximum violation is obtained by (4.8). 
The maximum violation of Bell inequality, $\eta_{\max}$, decreases monotonically with $s$. Table 2 sumarizes the results. We see that for all spins BI is broken. Note that there is a sharp decrease in $\eta_{\max}$ from $s = \fr{1}{2}$ to $s = 1$, while $\eta_{\max}$ decreases weakly as $s$ increases from 1. A possible reason is that, for $s = 1/2$ all states break BI while for $s \ge 1$ only a fraction of spin states break it.
\bc
{\bf Table 2} \\
\vspace{.1in} 

\bt{||c|c||c|c||c|c||}
\hline 
$s$ & $\eta_{\max} $ 
 & $s$ & $\eta_{\max}$  & $s$ & $\eta_{\max}$   \\
\hline 
$\fr{1}{2}$ & $\sq 2$ & $\fr{5}{2}$ & $1.1638$ &  $\fr{9}{2}$ & $1.1538$ \\
\hline 
 1 & $1.2112$  & 3 & $1.1599$ & 5 & $1.1526$ \\
\hline 
 $\fr{3}{2} $ & $1.1817 $ & $\fr{7}{2}$  & $1.1572$  & $\fr{11}{2}$ & $1.1517$ \\
\hline 
2 & $1.17$ & 4 & $1.1553$ & 6 & $1.1509$ \\
\hline
& & &  & $\iny$ & $1.143$  \\
\hline 
\et
\vspace{.1in}

Two successive measurements  

\ec

We now consider a case where the preparation of the pure state is noisy, resulting in a state 
$$ \rho(f) = \left( 1-f\right)  \rho_0^{\max} + \fr{f}{2s +1} I \eqno{(4.9)}$$ 
where the positive parameter $f \le 1$ is the probability of the noise contamination of the state.
Proceeding as before, we get 
$$ \lan \al_1 \al_2 \ran = \fr{1}{2} \cos \th_{12} [A' \cos^2 \th_{1} + B'] \eqno{(4.10)}$$ 
where 
$$ A' =(1-f)(2s-1)s; ~~~~ B' =(1-f) s + \fr{2}{3} f (s+1)s$$ 
which leads to 
$$ \eta_{noise} = \left( \fr{1}{2s^2}\right) (\sin \th_1 + \cos \th_1) (A' \cos^2 \th_1 + B') .\eqno{(4.11)}$$ 
Using the maximization procedure, $\th_1$ for maximum $\eta_{noise}$ is given by a real root of 
$$ B'\tan^3\th_1 + (2A'-B') \tan^2\th_1 + (3A' + B') \tan \th_1 - (A'+B') = 0. \eqno{(4.12)}$$ 
The range of $f$ for which $\eta_{noise} > 1$ is tabulated in Table 3. Note that for $s = \fr{1}{2}$ the state corresponding to $f=1$ (the random mixture) also breaks BI!  Of course we have already shown that for $s = \fr{1}{2}$ BI is broken for all states.

\bc
{\bf Table 3} \\

\vspace{.1in} 

\bt{||c|c||c|c||c|c||}
\hline 
$s$ & $f$ & $s$ & $f$ & $s$ & $f$ \\
\hline 
$\fr{1}{2}$ & $0 \le f \le 1$ & $\fr{5}{2}$ & $f < 0.287$ & $\fr{9}{2}$ & $f< 0.239$ \\
\hline 
 1 & $f<0.696$ & 3 & $f<0.267$ & 5 & $f<0.234$ \\
\hline 
$\fr{3}{2} $ & $f<0.395$ & $\fr{7}{2}$ & $f<0.254$ & $\fr{11}{2}$ & $f<0.230$\\
\hline 
2 & $f<0.321$  & 4 & $f<0.245$ & 6 & $f<0.227$ \\
\hline 
& & & & $\iny$ & $f<0.195$  \\
\hline 
\et
\vspace{.1in}

The range $f$ for the violation  of BI  
\ec

Table 3 answers the question, ``what is the maximum fraction of  noise that can be added to $ \rho_0^{\max}$, which maximally breaks BI, so that the state has stranger than classical correlations?'' we see that the corresponding fraction of noise decreases monotonically with s, or with the dimension of the Hilbert space. This can be compared with the results of Collins and Popescu[15] who find that the nonlocal character of the correlations between the outcomes of measurements performed on entangled systems separated in space is robust in the presence of noise. They show that, for any fraction of noise, by taking the Hilbert space of large enaugh dimension, we can find bipartite entangled states giving nonlocal correlations. These results are obtained by considering two successive measurements on each part of the system. In the present case of successive measurements on the single spin state, we see that the fraction of noise that can be added so that the quantum correlations continue to break Bell inequality, falls off monotonically with $s$, or the dimension of the Hilbert space. For $s=\frac{1}{2}$ all franctions $f \leq 1$ are allowed, while for large s, $f<0.195$.


\subsection{Three successive measurements (MKI)} 

We again assume the input state to be (4.1). Using Appendix A :
$$ \lan \al_1 \al_2 \al_3 \ran = \fr{1}{16}  \cos \th_{1} \cos \th_{23} [M \cos^2 \th_{12} + N] \eqno{(4.13)}$$ 
where 
 $$ M =\sum_{\al_0}p_{\al_0} [9 \al^3_0 + \al_0 (s(s+1) - 3)],~~~~~ 
 N = \sum_{\al_0}p_{\al_0}[ - 3\al^3_0 + \al_0 (5s(s+1) +1)] $$
Substitution in MKI and finding the conditions for which it is maximized, we get 
$ \th_1 = 0~,~ \th_{1'} = \fr{\pi}{2}~,~ \th_{3'} = \pi~,~ \th_2 + \th_{2'} = \pi$. 
Again we define $\eta=\left|MKI\right|/{s^3}$
$$ \eta = \left( \fr{1}{16s^3}\right) (\sin \th_2 + \cos \th_2) (M \cos^2 \th_2 + N) \eqno{(4.14)}$$ 
where $\th_2$ is real roots of 
$$ N \tan^3 \th_2 + (2M-N) \tan^2 \th_2 + (3M + N) \tan \th_2 - (M+N) = 0 \eqno{(4.15)}$$ 
Consider $s = \fr{1}{2}$.  In this case $M = 0$ and $N = 2(p_{1/2} -  p_{-1/2})$.  This gives $$ \eta_{\max} = \left| p_{1/2} - p_{-1/2} \right| \sq 2  \eqno{(4.16)}$$
 
For $p_{\al_0 = \fr{1}{2}} > 0.85$ and $p_{\al_0 = 1/2} < 0.15~~ \eta > 1$. Maximum violation $(\eta_{\max} = \sq 2)$ is obtained when one of $p_{1/2}, p_{-1/2}$ is zero, i.e. when the initial spin state is pure state. 

For spin 1 we get : $$ \eta_{\max} = (1.2112) | p_1 - p_{-1}| \eqno (4.17)\\$$

\noi for $\eta > 1 \Ra | p_1 - p_{-1}|  > 0.83.$
Maximum violation is 1.2112 and is obtained when $p_1 = 0$ or $p_{-1} = 0$ and $p_0 =0$ i.e. the input state is a pure state $| \vec{s} \cdot  \h a_0, + 1 \ran $ or $|\vec{s} \cdot  \h a_0,  -1 \ran$.

We now specialize to pure states of the form $\rho^{\max}_0 = | \vec{s} \cdot \h a_0, s \ran \lan \vec{s} \cdot \h a_0, s|$. 
Table 4 summarizes the results. We see that MKI is broken for $\fr{1}{2} \le s \le 3$ and for $s > 3$, it is satisfied.  
Since $\al_0 = s$ correspond to maximum $\eta$ for all states of spin $s$, we see that for $s > 3$, three successive measurements are classically correlated. 
It is straightforward to check that, three successive measurements satisfy  Svetlichny Inequality (SI). 
The reason is that, for all $s$, the settings of the measurement directions which maximize $MKI'$are obtained from those which maximize $MKI$ by interchanging primes on the corresponding unit vectors. Thus these two settings are incompatible so that we cannot get a single set of measurement directions, which maximize both $MKI$ and $MKI'$.  In fact, for all $s$, the measurement directions which maximize $MKI (MKI')$ correspond to $MKI' = 0 (MKI = 0)$. This result can be generalized to $n$ successive measurements on spin $\fr{1}{2}$ system.

\bc
{\bf Table 4} \\
\vspace{.1in} 

\bt{||c|c||c|c||c|c||}
\hline 
$s$ & $\eta_{\max} $ & $s$ & $\eta_{\max} $ & $s$ & $\eta_{\max} $ \\
\hline 
$\fr{1}{2}$ & $\sq 2$ &  $\fr{5}{2}$ & $1.0351$ & $\fr{9}{2}$ & $0.9666$ \\
\hline 
1 & $1.2112$ & 3 & $1.0103$ & 5 & $0.9575$ \\
\hline 
 $\fr{3}{2} $ & $1.1234 $  & $\fr{7}{2}$ & $0.9919$  & $\fr{11}{2}$ & $0.9499$\\
\hline 
 2 & $1.0702$  & 4 & $0.9778$  & 6 & $0.9436 $ \\
\hline 
& & & & $\iny$ & $0.87 $  \\
\hline 
\et

\vspace{.1in} 

Three successive  measurements

\ec

\section{ THE CASE OF $n$ SUCCESSIVE MEASUREMENTS}
\subsection{SPIN $s= \frac{1}{2}$}

We consider $n$ successive measurements in direction $\vec{s} \cdot \h a_i, (i = 1, 2, 3, \cdots, n)$ on a spin $s = \fr{1}{2}$ particle in mixed state. For simplicity we take the eigenvalues to be $\al_k = \pm 1$ i.e. eigenvalues of $\si_z$.  We also write $| \al_k \ran$ for $| s \cdot \h a_k, \al_k\ran$ 
$$ \rho_0 = p_+ | \al_0 = + \ran \lan \al_0 = +| + p_- |\al_0 = - \ran \lan \al_0 = - | \eqno{(5.1)}$$
For spin $\fr{1}{2}$ we have 
$$ |\lan \al_{k-1} | \al_k \ran |^2 = \fr{1}{2} (1 + \al_{k-1} \al_k \cos \th_{k-1,k}) \eqno{(5.2)}$$ 
$$ \cos \th_{k-1, k} = \h a_{k-1} \cdot \h a_k $$ 
so 
$$p(\al_1, \al_2, \cdots, \al_n) = \fr{1}{2^n} \prod^n_{i=1} (1 + \al_{i-1} \al_i \cos \th_{i-1, i}). \eqno{(5.3)}$$ 
For $n$ successive experiments on spin $\fr{1}{2}$ 
\benrr
\lan \al_{n-1} \al_n \ran & = & \sum_{\al_0 = \pm 1} p_{\al_0} \sum_{\al_1 \cdots \al_n = \pm 1} \al_{n-1} \al_n p(\al_1, \al_2, \cdots, \al_n)\\ 
& = & \sum_{\al_0 = \pm 1} p_{\al_0} 2^{-n} \prod^n_{i=1} \sum_{\al_i = \pm 1} \al_{n-1} \al_n (1 + \al_{i-1} \al_i \cos \th_{i, i-1})  =  \cos \th_{n-1, n} ~~~~~~~~~~\mbox{(5.4)}
\eenrr
\benrr
\mbox{Further}~~~\lan \al_n \ran & = & \sum_{\al_0 = \pm 1} p_{\al_0} \sum_{\al_1 \cdots \al_n} \al_n p(\al_1, \cdots, \al_n) \\
& = & \sum_{\al_0 = \pm 1} p_{\al_0} 2^{-n} \prod^n_{i=1} \sum_{\al_i = \pm 1} \al_n (1 + \al_{i-1} \al_i \cos \th_{i, i-1}) \\
& = & (p_+ - p_-) \cos \th_1 \cos \th_{12} \cdots \cos \th_{n-1,  n} ~~~~~~~~~~~~~~~~~~~~~~~~~~~~~~~~~~~~~~ \mbox{(5.5)}
\eenrr
(5.4) and (5.5) give: 
$$ \lan \al_n \ran = (p_+ - p_-) \lan \al_1 \ran \lan \al_2 \al_3 \ran \cdots \lan \al_{n-1} \al_n\ran .  \eqno{(5.6)}$$
Further 
$$ \lan \al_{n-k} \cdots \al_n \ran = \sum\limits_{\al_0} p_{\al_0} 2^{-n} \prod^n_{i=1} \sum\limits_{\al_i = \pm 1} (\al_{n-k} \cdots \al_n) (1 + \al_{i-1} \al_i \cos \th_{i-1,i})$$
$$ = \left\{ \ba{ll} (p_+ - p_-) \lan \al_1 \ran \lan \al_2 \al_3\ran \cdots \lan \al_{n-1} \al_n\ran & k ~~\mbox{even} \\\\
  \lan \al_{n-k} \al_{n-k+1}\ran \lan \al_{n-k+2}
 \al_{n-k+3} \ran \cdots \lan \al_{n-1} \al_n \ran & k ~~ \mbox{odd} \ea \right. \eqno{(5.7)}$$ \\

All of the above results are inherently quantum and are not compatible with HVT. The first two results ((5.5) and (5.6)) are the special cases of the last result for $k = 1$ and $k = 0$ (with $\al_0 = 1)$. If the number of variables (which are averaged) is odd (i.e. $k$ is even) the average depends on the measurements prior to $(n-k)$, while in the other case the average does not depend on the measurements prior to $(n-k)$. For example for two successive measurements, $k = 1$ gives $\lan \al_1 \al_2 \ran = \cos \th_{12}$ is independent of initial state. While for three expereiments $n = 3$ and $k = 2$ give $\lan \al_1 \al_2 \al_3 \ran = (p_+ - p_-) \lan \al_1 \ran \lan \al_2 \al_3 \ran$ showing its dependence on initial state. Interestingly if $\h a_0 \perp \h a_1$ so that $\lan \al_1 \ran = 0$ or the initial state is random $(p_+ = p_-)$ then for all even $k$, 
$ \lan \al_{n-k} \cdots \al_n \ran = 0$ or $ \lan \al_1 \al_2 \cdots \al_{n=2 p+1} \ran = 0. $ 
 


We shall now show that for $n$ successive experiments $(n > 1)$ QM violates$| \lan M_{n} \ran |$ upto $\sq 2$ for spin $\fr{1}{2} $. (We take the eigenvalues to be $\al_k = \pm 1$ so $ | \lan M_k \ran |_{HVT} \leq 1$) . We have already shown that for $n = 2$ and $n = 3$ (Section 4). 
Using equations (5.7) and (3.12) we find that 
$$ | \lan M_{k}\ran| = \fr{1}{2} | \lan M_{k-2}  \ran [\lan \al_{k-1} \al_{k}'\ran + \lan \al_{k-1}' \al_{k} \ran ] + \lan M_{k-2} '\ran [ \lan \al_{k-1} \al_{k} \ran - \lan \al_{k-1}' \al_{k}' \ran ] | \eqno{(5.8)}$$ 
so
$$ \max | \lan M_{k}\ran | =\frac{\sqrt{2}}{2} \{ | \lan M_{k-2}\ran |+ | \lan M'_{k-2}\ran | \} =\sq 2 \eqno{(5.9)}$$

Therefore, we conclude that QM violates $M_n$ inequality for $n$ successive measurements upto $ \sqrt{2}$ . 
\subsection{spin $s\geq1$}

In section 4 we studied BI and MKI for a class of mixed input quantum states for all spins. Section 5.1 dealt with $n$ successive measurements on a spin $1/2$ particle in a mixed state. Here we look at the effect of $n$ successive measurements on spin $s>\frac{1}{2}$ on a pure input state with $\alpha_{0}=s$, as we know that maximum violation is obtained for pure states with $\alpha_{0}=s$.

Consider $n$ successive measurements($n>3$) on spin s in state $\alpha_{0}=s$. The correlation $ \lan \al_{1}  \cdots , \al_n \ran $ of outputs $\alpha_{1},\cdots\alpha_{n}$ is given by
$$ \lan \al_{1} \cdots \al_n \ran= \sum^s_{\al_1 ,  \cdots , \al_n = -s} \al_1 p(\al_1, \cdots ,\al_n) \eqno{(5.10)}$$
By using (A.7), (A.11), (A.19) , we get
$$ \lan \al_{1} \cdots \al_n \ran=\fr{1}{16}  \cos \th_{n,n-1} \cos \th_{n-2,n-3} [M \cos^2 \th_{n-2,n-1} + N] \eqno{(5.11)} $$
where
$$N=-3 \lan \al_{1} \cdots \al^4_{n-3} \ran+[1+5s(s+1)] \lan \al_{1} \cdots \al^2_{n-3} \ran$$

$$M=9 \lan \al_{1} \cdots \al^4_{n-3} \ran+[-3+s(s+1)] \lan \al_{1} \cdots \al^2_{n-3} \ran$$

To obtain the maximum violation of $ | \lan M_n \ran |$ For $s>\frac{1}{2}$, the positive factor$N+M \cos^2 \th_{n-2,n-1}$ should have maximum value which is realized for $\al_{1} =\al_{2} =\cdots\al_{n-3} =s$ or, in other words, when all directions of quantization $\h a_{1} ,\h a_{2} ,\cdots ,\h a_{n-3}$ are parallel to $\h a_{0} $. This gives\\
$$ \lan \al_{1} \cdots \al_n \ran=\fr{s^{n-3}}{16} s \cos \th_{n,n-1} \cos \th_{n-2,n-3} [(10s^{2} +s-3) \cos^2 \th_{n-2,n-1} + (2s^{2} +5s+1)] \eqno{(5.12)} $$\\
Comparison with (4.13) gives
$$ \lan \al_{1} \cdots \al_n \ran=s^{n-3} \lan \al_{n-2} \al_{n-1} \al_n \ran\eqno{(5.13)}$$
Thus we see that the quantum correlation in $n$ successive measurements on the spin $s$ particle in $\al_{0} =s$ input state is proportional to the correlations in the last three measurements. To get the contact with HVT we use these correlations in the corresponding $|\lan M_{n} \ran |$. We get 

$$ \lan M_{n} \ran =\frac{1}{2}\lan M_{n-3} (M_{3} +M'_{3} )\ran + \frac{1}{2}\lan M'_{n-3} (M_{3} -M'_{3} )\ran \eqno{(5.14)} $$
since $\h a_{1} ,\h a_{2} ,\cdots ,\h a_{n-3}$ are parallel to  $\h a_{0} $  so $M_{n-3} =M'_{n-3} =s^{n-3} $, Therefore,

$$ \lan M_{n} \ran =\lan M_{n-3} M_{3}\ran = s^{n-3} \lan M_{3} \ran \eqno{(5.15)} $$
Defining $ \eta_{n} =\dfrac{| \lan M_{n} \ran |}{s^{n}} $, where the denominator gives the maximum value of $ M_{n}$ allowed by HVT, we get, using (5.15) :
$$  \eta_{n} =\dfrac{| \lan M_{3} \ran |}{s^{3}} = \eta_{3} \eqno{(5.16)}$$
Thus we see that for $ s> \frac{1}{2} $ , the violation of $ \lan M_{n} \ran $ by quantum correlations in $n$ successive measurements with $n>3$ is the same as $\eta_{3} $ and thus is independent of $n$. In section 4, we proved that $\eta>1$ for $s\leq3$ and $\eta<1$ for $s>3$. Thus quantum correlations break $ \lan M_{n} \ran $ for all $n>3$ when $s\leq3$ and satisfy $ \lan M_{n} \ran$ for all $n>3$ when $s>3$.\\

 We have confirmed this result numerically upto $n=5$ for $s \leq  3$. In confirmity with section 4, we found that all successive measurements break MKI with same value of $\eta$ for $s=\frac{1}{2} $ and $s=1$.

\section{CLASSICAL SIMULATION OF $n$ SUCCESSIVE MEASUREMENTS ON A SPIN $\fr{1}{2}$ SYSTEM}

We have seen that QM correlations between the outputs of $n$ successive measurements of incompatible observables $\vec{s} \cdot \h a_k (k = 1, 2, \cdots n)$ are stronger than their classical (HVT) counterparts. An interesting question is whether these quantum correlations can be simulated classically?  Can we design a classical protocol to produce $n$ sets of outputs which are correlated as if these were the outputs of genunine quantum measurements?  If this is possible, what amout of classical information (cbits) has to be shared between successive measurements? [16]  We try and answer some aspects of these questions in this section. Notice that, there is no room for non-locality in this scenario, because the events are time-like separated. When the particle is coming out from $i$-th experiment there is no particle in any of the subsequent experiments. The communication of information is done by the particle itself. We now describe our protocol for two successive measurements. 

We imagine that two experimenters, Alice and Bob perform two successive measurements of $\vec{s} \cdot \h a_1$ and $\vec{s} \cdot \h a_2$.  Directions $\h a_1$ and $\h a_2$ are chosen by each experimenter randomly and independent of each other. Alice and Bob do not know each others inputs $(\h a_1, \h a_2)$ and outputs $(\al_1, \al_2)$. Alice knows the input state parameter $\h a_0$. Bob does not know $\h a_0$. They share three random variables (unit vectors) $\h \la_0, \h \la_1, \h \la_2$. They are chosen independently and distributed uniformly over the unit sphere. The protocol proceeds as follows: (i) Alice outputs $\al_1 = sgn[\h a_1 \cdot (\h \la_0 + \h a_0)]$. (ii) Alice sends two cbits $c_1$ and $c_2 \in \{ -1, 1\}$ to Bob where $c_1 = sgn[\h a_1 \cdot (\h \la_0 + \h a_0)] sgn(\h a_1 \cdot \h \la_1) = \al_1 ~ sgn(\h a_1 \cdot \h \la_1), c_2 = sgn[\h a_1 \cdot (\h \la_0 + \h a_0)] sgn(\h a_1 \cdot \h \la_2) = \al_1 ~ sgn(\h{a_1} \cdot \h{\la_2})$.  (iii) Bob outputs 
 $\al_2 = sgn[\h a_2 \cdot (c_1 \h \la_1 + c_2 \h \la_2)]$, where we have used the sgn function defined by $sgn(x) = +1$ if $x \ge 0$ and $sgn(x) = -1$ if $x < 0$.  We note immediately that Bob cannot obtain any information about Alice's input and output from $c_1$ and $c_2$. We now show that the above protocol reproduces the statistics of two successive measurements of $\vec{s} \cdot \h a_1$ and $\vec{s} \cdot \h a_2$ on spin $1/2$ particle in initial state $|\vec{s} \cdot \h a_0, + \ran \lan \vec{s} \cdot \h a_0, + |$. As shown in Appendix B we have 
$$ \lan \al_1\ran = \h a_0 \cdot \h a_1~,~  \lan \al_1 \al_2 \ran = \h a_1 \cdot \h a_2~,~
 \lan \al_2\ran = (\h a_0, \h a_1) (\h a_1, \h a_2) = \lan \al_1 \ran \lan \al_1 \al_2 \ran$$ 
which is consistent with the quantum case. We can generalize this protocol to get the clasical simulation of $n$ successive experiments. Here, again, each experiment is performed by an independent experimenter, who has no knowledge of the inputs and outputs of the previous and the future experiments. All experimenters share $(2n+1)$ random variables (unit vectors) $\h \la_0, \h \la_1, \h \la_2, \cdots, \h \la_{2n}$. The $i$-th experimentor $(i > 1)$ receives cbit $c_{2i-3}$ and $c_{2i-2}$ from $(i-1)$-th experiment, defined by
$c_{2i-3} = \al_{i-1} sgn(\h a_{i-1} \cdot \h \la_{2i-3}),~~ c_{2i-2} = \al_{i-1} sgn(\h a_{i-1} \cdot \h \la_{2i-2}).$
The $i$-th experimenter, then outputs 
$ \al_i = sgn[\h a_i \cdot (c_{2i-3} \h \la_{2i-3} + c_{2i-2} \h \la_{2i-2})]$.
For $i = 1$, the outputs $\al_1 = sgn[\h a_1 \cdot (\h \la_0 + \h a_2)].$ \\

As shown in Appendix B, this protocol produces all quantum correlations between $n$ successive measurements ((5.4), (5.5), (5.6) and (5.7)).

\section{SUMMARY AND COMMENTS} 

In this paper we present a new approach for testing whether QM is consistant with the realistic property of a SHVT. In all the previous scenarioes comparing HVT and QM the principal hypothesis being tested was that in a given state HVT implies the existance of a joint probability distribution for all observables even if some of them are not compatible. QM is shown to contradict the consequence of this requirement as it does not assign joint probabilities to the values of incompatible observables. The particular implication that is tested is whether the marginal of the observable A in the joint distribution of the compatible observables A and B is the same as the marginal for A in joint distribution for the observables A and C even if B and C are not compatible. In other words, HVT implies noncontextuality for which QM can be tested. The celebrated theorem of Bell and Kochen-Specker showed that QM is contextual[17,18]. In our scenario the set of measured observables have a well defined joint probability distribution as each of them acts on a different state. Note that the Bell type  inequalities we have derived follow from equation (3.4) which says that, for a given value of stochastic hidden variable $\lambda$ the joint probability for the outcomes of successive measurements must be statistically independent. In other words the hidden variable $\lambda$ completly decides the probabilities of individual measurement outcomes independent of other measurements. We show that QM is not consistant with this requirement of HVT. A Bell type inequality, testing contextuality of QM was proposed by S.Basu, S.Bandyopadhyay, G.Kar and D.Home[19] and was shown that it could be emprically tested. However, the approach given in the present paper furnishes a test for realistic nature of QM independent of contextuality.
One advantage of this approach is that it can be used to get a measure of the deviation of QM from HVT. One such measure is the amount of information needed to be transferred between successive measurements in order to classically simulate quantum correlations. As we have shown in section 6, a pure spin 1/2 state can be classically simulated by communicating two c-bits of information to get the $k$-th  output from $(k-1)$-th  output by using (2k+1) shared random variables. Whether this is the minimum communication required is still open. For our protocol, the amount of information needed is twice as much in the case of bipartite nonlocal scenario[16].

 In sections 4 and 5 we have studied QM from HVT for different values of spin and for different number of successive measurements. The dependence of the deviation of QM from HVT on the spin value and on the number of successive measurements opens up new possibilities for comparison of these models, and may lead to a sharper understanding of QM.  We get many surprising results. First, for a spin $s$ particle, maximum deviation ($\eta$) is obtained for all convex combinations (mixed states) of $\al_0 = \pm s$  states. This is surprising as one would expect pure states to be more `quantum'  than the mixed ones thus breaking Bell inequalities by larger amount. In particular, all spin$1/2$ states maximally break Bell inequality as against only the entangled states break it in bipartite case. This does not contradict Bell's explicit construction of HVT for spin $1/2$ particle, as it does not apply for two or more successive measurements. Further, the maximum deviation from Bell inequality measured by $\eta_{max}$ falls off as the spin of the particle increase. There is a large drop in $\eta_{max}$ value from $s=1/2$ to $s=1$, after which it drops monotonically with $s$, but very weakly asymptotically approaching $\eta=1.4$. This can be compared with the case of two spin $s$ operators in the singlet state where the deviation from Bell's inequality is found to tend to a constant [20,21,22]. Three successive measurements violate MKI upto $s=3$. For $s>3$  MKI is satisfied by all states. All spins satisfy SI for three successive measurements. In section 5 we show that for fixed $S\leq 3$, $n$ successive measurements break all the MK inequalities, and this is independent of  $n$, (excepting, a small drop in maximum $\eta$ value from $n=2$ to $n=3$ for $s\leq 1$.) In the case of $n$ spin 1/2 particles in the singlet state, Bell inequalities are broken by a factor which increases exponentially with $n$ [22].

 As a final remark, it would be interesting to consider Bell inequalities involving both two and three successive measurements correlations. A straightforward calculation would allow us to prove that HVT satisfy the following inequality [23]: 
$$-5\leq \lan \al_1 \al_2 \al_3'\ran - \lan \al_1 \al_2' \al_3' \ran - \lan \al_1' \al_2 \al_3' \ran - \lan \al_1' \al_2' \al_3\ran-\lan \al_1 \al_2 '\ran - \lan \al_1 \al_3' \ran - \lan  \al_2 \al_3 \ran \leq 3 $$
$$ -8 \leq \lan \al_1 \al_2 \al_3'\ran - \lan \al_1 \al_2' \al_3' \ran - \lan \al_1' \al_2 \al_3' \ran - \lan \al_1' \al_2' \al_3\ran-2\lan \al_1 \al_2 '\ran -2 \lan \al_1 \al_3' \ran - 2\lan  \al_2 \al_3 \ran \leq 4 $$
It is not difficult to show that three successive measurements correlations for spin 1/2 break the hybrid Bell inequalities. So two successive measurements correlations are relevant to those of three successive mrasurements. This behaviour is analogous to three particle W state. We note that three particle GHZ state does not break the second inequality [23].

\begin{center}
\large {\bf ACKNOWLEDGEMENTS}
\end{center}
It is a pleasure to acknowledge Guruprasad Kar, Sibasish Ghosh, Debasish Sarkar, Samir Kenkuri for useful discussions.\\


\bc

{\large {\bf APPENDIX A}} 
\ec 
We evaluate $ \lan \al_1 \ran $, $ \lan \al_1 \al_2 \ran $ and $ \lan \al_1 \al_2 \al_3 \ran $ in the state $ \rho_0 $ given in(4.1).\\$(|\vec{s} \cdot \h a_0, \al_0 \ran \equiv | \h a_0, \al_0 \ran )$ 
$$
 \lan \al_1 \ran = \sum^s_{\al_1 = -s} \al_1 p(\al_1) = \lan \h a_0, \al_0 |\vec{s} \cdot  \h a_1 | \h a_0, \al_0\ran  =   \lan \h a_1, \al_0 | e^{i\vec{s}\cdot \h n \th_{1}} (\vec{s} \cdot \h a_1) e^{-i\vec{s} \cdot \h n \th_{1}}| \h a_1, \al_0 \ran \eqno{(A.1)}$$
where $\th_{1}$ is the angle between $\h a_0$ and $\h a_1$ and $\h n$ is the unit vector along the direction defined by $\h n = \h a_0 \times \h a_1$. 
By using Baker- Hausdorff Lemma 
$$ e^{iG\la} Ae^{-iG\la} = A + i\la [G, A] + \left( \fr{i^2 \la^2}{2!}\right) [G, [G, A]] + \cdots \eqno{(A.2)}$$ 
We get 
\benrr
\lan \al_1\ran  &= &  \lan \h a_1, \al_0 | \vec{s} \cdot \h a_1 | \h a_1, \al_0 \ran + \fr{i\th_{1}}{1!} \lan \h a_1, \al_0| [\vec{s} \cdot \h n, \vec{s} \cdot \h a_1] | \h a_1, \al_0 \ran \\
& &  + \fr{i^2 \th^2_{1}}{2!} \lan \h a_1, \al_0 | [ \vec{s} \cdot \h n, [ \vec{s} \cdot \h n, \vec{s} \cdot \h a_1]] | \h a_1 \al_0 \ran + \cdots~~~~~~~~~~~~~~~~~~~~~~~~~~~~~~~~~~~~~~~ \mbox{(A.3)}
\eenrr
$$ \lan \h a_1, \al_0 | \vec{s} \cdot \h a_1 | \h a_1, \al_0 \ran = \al_0 \eqno{(A.4)}$$ 

$$ \lan \h a_1, \al_0 | [\vec{s} \cdot \h n, \vec{s} \cdot \h a_1] |\h a_1, \al_0 \ran = \lan \h a_1, \al_0 | (i\vec{s} \cdot (\h n \times \h a_1)) | \h a_1, \al_0 \ran = 0 \eqno{(A.5)}$$ 

$$ \lan \h a_1, \al_0 | [\vec{s} \cdot \h n, [\vec{s} \cdot \h n, \vec{s} \cdot \h a_1]] |\h a_1, \al_0 \ran = \lan \h a_1, \al_0 | \vec{s} \cdot \h a_1 |\h a_1, \al_0 \ran = \al_0\eqno{(A.6)}$$ 
Terms with odd powers of $\th_{1}$ vanish 
$$ \lan  \al_1 \ran = \al_0 - \fr{\th^2_{1}}{2!} \al_0 + \fr{\th^4_{1}}{4!} \al_0 - \cdots = \al_0 \cos \th_{1} \eqno{(A.7)}$$ 
If the initial state is mixed state(4.1):
$$ \lan  \al_1 \ran =  \sum^{+s}_{\al_0= - s}p_{\al_0}\al_0 \cos \th_{1} \eqno{(A.8)}$$ 
Further we compute 
$$ \lan \al_1 \al_2 \ran = \sum_{\al_1} \al_1 | \lan \h a_0, \al_0 | \h a_1, \al_1 \ran |^2 \sum_{\al_2}\al_2 | \lan \h a_1, \al_1 | \h a_2, \al_2 \ran |^2$$ 
By using (A.7) 
\begin{eqnarray*}
\lan \al_1 \al_2 \ran & = & \cos \th_{12} \sum_{\al_1} \al^2_1 | \lan \h a_0, \al_0 |\h a_1, \al_1 \ran |^2 
 =  \cos \th_{12} \lan \h a_0, \al_0 | (\vec{s} \cdot \h a_1)^2| \h a_0, \al_0 \ran \\
& =& \cos \th_{12}   \lan \h a_1, \al_0 | e^{i\vec{s}\cdot \hat{n} \th_{1}} (\vec{s} \cdot \h a_1)^2 e^{-i\vec{s} \cdot \hat{n}\th_{1}} | \h a_1, \al_0 \ran ~~~~~~~~~~~~~~~~~~~~~~~~~~~~~~~~~~~~~~~~(A.9)
\end{eqnarray*}
Using the Baker-Hausdorff Lemma, and using 
$$\lan \hat a_{1} ,\al_{0} | [\vec{s} \cdot \h n, [ \vec{s} \cdot \h n, [ \vec{s} \cdot \h n, \cdots [\vec{s} \cdot \hat{n}, (\vec{s} \cdot \hat{a_1})^2]] \cdots]]| \hat a_{1} ,\al_{0}\ran \eqno{(A.10)}$$ 
$$ = \left\{ \ba{l} 0~~~~~~~~~~~~~~~~~~~~~\mbox{if $\vec{s} \cdot \h n$ occurs odd number of times } \\ \\ 3\al^2_0 - s^2 - s~~~~~~~\mbox{if $\vec{s} \cdot \h n$ occurs $2p$ times} \ea \right. $$
We get 
$$\lan  \al_1 \al_2  \ran = \fr{1}{2} \cos \th_{12} [(s^2 + s - \al^2_0) + (3\al^2_0 - s^2 - s) \cos^2 \th_{1}] \eqno{(A.11)}$$ 
If the initial state is mixed state(4.1):
$$\lan  \al_1 \al_2  \ran = \fr{1}{2} \cos \th_{12}  \sum^{+s}_{\al_0= - s}p_{\al_0}[(s^2 + s - \al^2_0) + (3\al^2_0 - s^2 - s) \cos^2 \th_{1}] \eqno{(A.12)}$$ 
Next we calculate : 
$$ \lan \al_1 \al_2 \al_3 \ran = \sum_{\al_1} \al_1 | \lan \h a_0,  \al_0 | \h a_1, \al_1 \ran |^2 \sum_{\al_2} \al_2 | \lan \hat{a_1}, \al_1 | \h a_2, \al_2 |^2 \sum_{\al_3} \al_3 | \lan \h a_2, \al_2 | \h a_3, \al_3 \ran |^2$$
By using (A.7) 
$$ = \cos \th_{23} \sum_{\al_1} \al_1 |\lan \h a_0, \al_0 |\h a_1, \al_1 \ran |^2 \sum_{\al_2} \al^2_2 | \lan \h a_1, \al_1 | \h a_2, \al_2 \ran|^2.$$
By using (A.11) 
$$ = \cos \th_{23} \sum_{\al_1} \al_1 |\lan \h a_0, \al_0 | \h a_1, \al_1 \ran |^2  \fr{1}{2} \{ (s^2 +s - \al_1)^2 + (3\al^2_1 - s^2 - s)  \cos^2 \th_{12} \} $$ 
This simplifies to 
$$ \lan \al_1 \al_2 \al_3 \ran = \fr{1}{2} \al_0 \cos \th_{1} \cos \th_{23} \sin^2 \th_{12} s(s+1) + \fr{1}{2} \cos \th_{23} (3 \cos^2 \th_{12} - 1) A \eqno{(A.13)}$$ 
where 
$$ A = \sum_{\al_1} \al^3_1 | \lan \h a_0,  \al_0 |\h a_1, \al_1 \ran |^2 = \lan \h a_1, \al_0| e^{i\vec{s}\cdot \hat{n}\th_{1}} (\vec{s} \cdot \h a_1)^3 e^{-i\vec{s}\cdot \hat{n}\th_{1}}|\h a_1, \al_0\ran \eqno{(A.14)}$$ 
Using Baker-Hausdorff lemma and 
$$\lan \hat a_{1} ,\al_{0} | [\vec{s} \cdot \h n, [ \vec{s} \cdot \h n, [ \vec{s} \cdot \h n, \cdots [\vec{s} \cdot \hat{n}, (\vec{s} \cdot \hat{a_1})^3]] \cdots]]| \hat a_{1} ,\al_{0}\ran \eqno{(A.15)}$$ 
$$ = \left\{ \ba{l} 0~~~~~~~~~~~~~~~~~~~~~\mbox{if $\vec{s} \cdot \h n$ occurs odd number of times } \\ \\ Y (X - a^3_0) + X ~~ \mbox{if $\vec{s} \cdot \h n$ occurs $2p$ times} \ea \right. $$
 

$$ X = 6\al_0^3 + \al_0 (1 - 3s(s+1))$$ 
$$Y=3^{2p-2} + 3^{2p-4} + \cdots + 3^2$$

We get

$$A = \al^3_0 + X(\cos \th_{1} - 1) + (X - \al_0^3) f(\th_{1})\eqno{(A.16)}$$ 

 where 
$$ f(\th) = \fr{\th^4}{4!} 3^2 - \fr{\th^6}{6!} (3^2 + 3^4) + \fr{\th^8}{8!} (3^2 + 3^4 + 3^6) + \cdots .$$ 

This function satisfies $ f''(\th) + 9f(\th) = 9 - 9 \cos \th $ 
whose general solution is 
$$ f(\th) = 1 - \fr{9}{8} \cos \th \eqno{(A.17)}$$ 
This gives 
 $$ A = \fr{1}{8} \cos \th_{1} (3\al^3_0 + 3\al_0 s(s+1) - \al_0) \eqno{(A.18)}$$ 
After substituting (A.18) in (A.13) and simplify : 
$$ \lan \al_1 \al_2 \al_3 \ran = \fr{1}{16} \cos \th_{1} \cos \th_{23} [M \cos^2 \th_{12} + N] \eqno{(A.19)}$$ 
$$M = \al_0[9 \al^2_0 + s(s+1) - 3],~~~ N = \al_0[ - 3\al^2_0 + 5s(s + 1) + 1] .$$ 
If the initial state is mixed state(4.1):
$$ M = \sum^{+s}_{\al_0= - s}p_{\al_0} \al_0[9 \al^2_0 + s(s+1) - 3], ~~~ N =  \sum^{+s}_{\al_0= - s}p_{\al_0}\al_0[ - 3\al^2_0 + 5s(s + 1) + 1] \eqno{(A.20)} $$ 





 


\bc
{\large {\bf APPENDIX B}} \\
\ec 

To evaluate $\lan \al_1 \ran = \h a_1 \cdot \h a_0$ we integrate over $\h \la_0$, taking $\h a_1$ to point along the positive $\h z$ axis. 
\benrr
\lan \al_1 \ran & = & \fr{1}{4\pi} \int d\la_0 sgn[\h a_1 \cdot (\h \la_0 + \h a_0)] \\
& = & \fr{1}{4\pi} \int^{2\pi}_0 d\b_0 \int^\pi_0 \sin~ \al_0 d \al_0 ~ sgn(cos \al_0 + \cos \th_1) = \cos \th_1 = \h a_1 \cdot \h a_0 ~~~~~ \mbox{(B-1)}
\eenrr
where $\cos \al_0 = \h a_1 \cdot \h \la_0$ and $\h \la_0 = (\sin \al_0  \cos \b_0, \sin \al_0 \sin \b_0, \cos \al_0)$.

To evaluate $\lan \al_2 \ran = (\h a_0 \cdot \h a_1) (\h a_1 \cdot \h a_2)$ 
\benrr
\lan \al_2 \ran & = & \fr{1}{(4\pi)^3} \int d\la_0 d\la_1 d\la_2 ~ sgn[\h{a_2}\cdot (c_1\h\la_1 + c_2\h\la_2)] \\
& = & \fr{1}{(4\pi)^3} \int d\la_0d\la_1 d\la_2 \fr{1}{4} \sum_{d_1 =\pm1} \sum_{d_2=\pm 1} (1+c_1d_1) (1+c_2d_2) sgn[\h a_2 \cdot (d_1\h \la_1 + d_2\h \la_2)]\\
& = & \fr{1}{(4\pi)^3} \fr{1}{2} \int d\la_0 d \la_1 d\la_2 \{c_1(sgn[\h a_2 \cdot (\h \la_1 + \h \la_2)] + sgn[\h a_2 \cdot (\h \la_1 - \h \la_2)]) \\
& & + c_2(sgn[\h a_2 \cdot (\h \la_1 + \h \la_2)] - sgn[\h a_2 \cdot (\h \la_1 - \h \la_2)])\} \\
& = & \fr{1}{(4\pi)^2} \fr{1}{2} \int d\la_0 sgn[\h a_1 \cdot (\h \la_0 + \h a_0)] 
 \{\int d\la_1 sgn(\h a_1 \cdot \h \la_1) 2(\h a_2 \cdot \h \la_1)\\
& &  + \int d\la_2 sgn(\h a_1 \cdot \h \la_2) 2(\h a_2 \cdot \h \la_2)\} 
 =  (\h a_0 \cdot \h a_1) (\h a_1 \cdot \h a_2) ~~~~~~~~~~~~~~~~~~~~~~~\mbox{(B-2) }
\eenrr
The same way we can prove 
$$ \lan \al_1 \al_2 \ran = \fr{1}{(4\pi)^3} \int d \la_0 d \la_1 d\la_2 \al_1 \al_2 = (\h a_1 \cdot \h a_2) \eqno{(B-3)}$$ 
By using induction, we shall show that for $n(n > 2)$ successive measurements is simulated by this protocol.

We suppsoe for $n = k-1$, it is true i.e. 
$$ \lan \al_{k-1}\ran = \fr{1}{(4\pi)^{2k-4}} \int d\la_0 d\la_1 \cdots d\la_{2k-4} \al_{k-1} = \lan \al_1\ran \lan \al_2\al_3 \ran \cdots \lan \al_{k-2} \al_{k-1}\ran \eqno{(B-4)} $$
$$ \lan \al_{k-2} \al_{k-1} \ran = \fr{1}{(4\pi)^{2k-4}} \int d\la_0 d\la_1 \cdots d\la_{2k-4} \al_{k-2} \al_{k-1}  = \h a_{k-2} \cdot \h a_{k-1} \eqno{(B-5)}$$ 

$$ \lan \al_{k-1-m} \cdots \al_{k-1}\ran = \left\{ \ba{ll} \lan \al_1 \ran \lan \al_2 \al_3 \ran \cdots \lan \al_{k-2} \al_{k-1} \ran & m~~ \mbox{even} \\ \\
\lan \al_{k-1-m} \al_{k-m} \ran \cdots \lan \al_{k-2} \al_{k-1} \ran & m~~ \mbox{odd} \ea \right. \eqno{(B-6)}$$

So, for $n = k$, first we show that, 
\benrr
& & \int d\la_{2k-3} d\la_{2k-2} \al_k = \int d\la_{2k-3} d\la_{2k-2} sgn[\h a_k \cdot (c_{2k-3} \h \la_{2k-3} + c_{2k-2} \h \la_{2k-2})]\\
& & = \fr{1}{2} \int d\la_{2k-3} c_{2k-3} \int d\la_{2k-2} [sgn(\h a_k \cdot (\h \la_{2k-3} + \h \la_{2k-2})) + sgn(\h a_k \cdot (\h \la_{2k-3} - \h \la_{2k-2}))]\\
& & + \fr{1}{2} \int d\la_{2k-2} c_{2k-2} \int d\la_{2k-3} [sgn(\h a_k \cdot (\h \la_{2k-3} + \h \la_{2k-2})) - sgn(\h a_k \cdot (\h \la_{2k-3} - \h \la_{2k-2}))]\\
& & = \int d\la_{2k-3} c_{2k-3} (4\pi) (\h a_k \cdot \h \la_{2k-3}) + \int d\la_{2k-2} c_{2k-2} (4\pi) (\h a_k \cdot \h \la_{2k-2}) \\
& & = (4\pi) \al_{k-1} \{ \int d\la_{2k-3} sgn(\h a_{k-1} \cdot \h \la_{2k-3})(\h a_k \cdot \h \la_{2k-3}) + \int d\la_{2k-2} sgn(\h a_{k-1} \cdot \h \la_{2k-2}) (\h a_k \cdot \h \la_{2k-2})\}\\
& & = (4\pi)^2 \al_{k-1} \{ \fr{1}{2} (\h a_{k-1} \cdot \h a_k) + \fr{1}{2} (\h a_{k-1} \cdot \h a_k)\} = (4\pi)^2 \al_{k-1} (\h a_{k-1} \cdot \h a_k) ~~~~~~\mbox{(B-7)}
\eenrr 

By using (B-7), (B-4), (B-5) and (B-6) all quantum correlations are obtained by this protocol. 

\newpage

\bc
{\bf REFRENCES} \\
\ec 

\begin{verse} 

[1] Asher Peres, {\it Quantum Theory : Concepts and Methods}, (Kluwer Academic Publishers 1993).\\ 

[2] S.Popescu ,Phys. Rev. Lett.\textbf{ 74}, 14 (1995).\\

[3] A.J.Leggett and A.Garg, Phys.Rev.Lett.\textbf{ 54}, 857 (1985).\\

[4] O.Alter and Y.Yamamoto, \textit{Quantum Measurement of a Single System},(A Wiley-interscience Publication John Wiley and sons,INC)\\

[5]  A. Fine, Phys.Rev.Lett. \textbf{48}, 291 (1982).\\

[6] E. Anderson, S. M. Barnett and A. Aspect, Phys. Rev.A \textbf{72}, 042104 (2005).\\ 

[7] L. E. Ballentine, {\it Quantum Mechanics}, (prentice Hall, Englewood Cliffs, NJ, 1990). \\

[8] Michael Redhead, {\it Incompleteness, Nonlocality, and Realism}, (Clarendon Press, Oxford : 1987).\\ 

[9] A. Shimony, In Sixty-Two Years of Uncertainty. {\it Historical Philosophica, and Physical Inquiries into the Foundations of Quantum Mechanics,} edited by A. Miller (Plenum, New York, 1990), pp. 33-43.\\ 

[10] J. P. Jarrett (1984), On the physical significance of the locality conditions in the Bell Arguments', Nous 18, 569-89.\\

[11] G. Svetlichny, Phys.Rev.D, \textbf{35} , 3066 (1887). \\

[12] M. Seevinck and G. Svetlichny, Phys.Rev.lett.\textbf{89},060401 (2002).\\

[13] D. Collins, N. Gisin, S. Popescu, D. Roberts, V. Scarani. Phys. Rev. Lett.\textbf{ 88 }, 170405(2002).\\

[14] R.F.Werner,Phys.Rev.A \textbf{40}4277(1989).\\

[15] D.Collins, S.Popescu, J.Phys.A :Math.Gen \textbf{34} no 35,6821(2001).\\

[16] B. F. Toner and O. Bacon, Phys. Rev. Lett.\textbf{ 91},187904 (2003). \\

[17] J.S.Bell, Rev.Mod.Phys.\textbf{38},447(1966).\\

[18] S.Kochen and E.P.Specker,  J.Math.Mech.\textbf{17}, 59 (1967).\\

[19] S.Basu, S.Bandyopadhay, G.Kar and D.Home, Phys.Lett.A,\textbf{279},284(2001).\\

[20] A.Peres, Phys.Rev.A\textbf{46},4413(1992).\\

[21] N.Gisin and A.Peres, Phys.Lett.A\textbf{162},15 (1992).\\

[22] A.Cabello, Phys. Rev. A \textbf{65}  062105 (2002).\\

[23] A.Cabello, Phys. Rev. A \textbf{65}  032108 (2002).\\

\end{verse}

\end{document}